\documentstyle[12pt]{article}
\input{epsf.tex}
\date{today}
\title{\bf Emission of intermediate mass fragments from hot $^{116}$Ba$^*$\\
formed in low-energy $^{58}$Ni+$^{58}$Ni reaction}
\author{M. Balasubramaniam, Rajesh Kumar and Raj K. Gupta\\
{\it Physics Department, Panjab University, Chandigarh-160014, India.}\\
C. Beck\\
{\it Institut de Recherches Subatomiques, UMR7500, IN2P3/ Universit\'e}\\
{\it Louis Pasteur, B.P. 28, F-67037 Strasbourg Cedex 2, France.}\\
Werner Scheid\\
{\it Institut f\"ur Theoretische Physik, Justus-Liebig-Universit\"at,}\\
{\it Heinrich-Buff-Ring 16, D-35392, Giessen, Germany.}\\
}
\date{\today}
\begin{document}
\maketitle
\def\be{\begin{equation}}
\def\ee{\end{equation}}
\def\ba{\begin{array}}
\def\ea{\end{array}}
\def\bea{\begin{eqnarray}}
\def\eea{\end{eqnarray}}
\begin{abstract}
The complex fragments (or intermediate mass fragments) observed in the low-energy 
$^{58}$Ni+$^{58}$Ni$\rightarrow ^{116}$Ba$^*$ reaction, are studied within the 
dynamical cluster decay model for s-wave with the use of the temperature-dependent 
liquid drop, Coulomb and proximity energies. The important result is that, due to 
the temperature effects in liquid drop energy, the explicit preference for 
$\alpha$-like fragments is washed out, though the $^{12}$C (or the complementary 
$^{104}$Sn) decay is still predicted to be one of the most probable 
$\alpha$-nucleus decay for this reaction. The production rates for non-$\alpha$
like intermediate mass fragments (IMFs) are now higher and the light particle 
production is shown to accompany the IMFs at all incident energies, without 
involving any statistical evaporation process in the model. The comparisons 
between the experimental data and the (s-wave) calculations for IMFs production 
cross sections are rather satisfactory and the contributions from other 
$\ell$-waves need to be added for a further improvement of these comparisons 
and for calculations of the total kinetic energies of fragments. 
\end{abstract}

Pacs No. {25.70.Jj, 23.60.+e, 23.70.+j, 24.10.-i} 
\newpage
\section{INTRODUCTION}
In low-energy heavy-ion reactions, intermediate mass fragments (IMFs) i.e. complex 
fragments with masses lighter than A$<$20, are emitted from both the light 
(A$\sim$60) and medium-mass (A$\sim$110) compound systems, and may arise as multiple 
"clusters", and are accompanied by multiple light particles (Z$\le$2) production. 
This light-particle emission, associated with evaporation residues production, is 
very satisfactorily understood as the fusion-evaporation process from an 
equiliberated compound nucleus in the statistical Hauser-Feshbach formalism
\cite{charity88,campo91,sanders89,sanders91,matsuse97}, described in evaporation 
codes such as LILITA and CASCADE or many other codes. The Hauser-Feshbach formalism 
has been extended to include the above noted observed complex IMFs, for example in 
the BUSCO code \cite{campo91} or in the Extended Hauser-Feshbach Method based on 
the scission-point picture \cite{matsuse97}. An alternative process for the IMFs 
production is the binary decay of a compound nucleus in the statistical fission 
model of Moretto \cite{moretto75} or of Vandenbosch and Huizenga \cite{vandenbosch73}. 
The fission model of \cite{moretto75} is used in the GEMINI code \cite{charity88} 
for the heavier compound systems (A$>$100), whereas the saddle-point transition-state 
model based on fission model of \cite{vandenbosch73} is used for the lighter systems 
(A$<$80) \cite{sanders91}. The emission of light particles (n, p, $\alpha$ or 
$\gamma$-ray) in both of these fission models is still calculated within the 
Hauser-Feshbach formalism \cite{sanders99}.

We know that the fission and cluster-decay are competing processes with the latter 
being more prevelent for lighter fragments \cite{gupta94}. In view of this result, 
in a very recent paper \cite{gupta02}, we have proposed for the first time an 
alternative new "cluster decay" process for the IMFs emitted from an excited 
$^{116}$Ba$^*$ compound nucleus produced in the low-energy $^{58}$Ni+$^{58}$Ni 
reaction. The IMFs are shown to arise as multiple "clusters" of masses A$<$20, and 
only at E$_{lab}>$200 MeV, in agreement with experimental results of Beckerman et al. 
\cite{beckerman81}. This result has its origin in the macroscopic liquid drop energy, 
since the shell effects are almost zero at the excitation energies involved. The 
multiple light particle (Z$\le$2) emission, other than via the statistical 
evaporation process (not treated in this model), is also shown to become equally 
probable at higher energies where only the pure liquid drop energies enter the 
calculations. The macroscopic liquid drop energy was, however, taken to be
temperature-independent \cite{gupta02} and only the angular momentum $\ell =0$ case 
was considered. Also, a generalized liquid-drop model has been proposed recently by
Royer and Moustabchir \cite{royer01} to describe the light-nucleus emission.

In this paper, we extend our previous work \cite{gupta02} by using the
temperature-dependent liquid drop energy (V$_{LDM}$(T)) from Davidson et al.
\cite{davidson94}, since the compound nucleus is hot enough at the excitation 
energies of interest. The other terms of the potential, that constitute the 
scattering potential V(R), are also taken temperature-dependent, based on Refs. 
\cite{davidson94,royer92}. In other words, we use a complete temperature-dependent 
potential and apply it to the medium-mass, hot compound system $^{116}$Ba$^*$ 
formed in $^{58}$Ni+$^{58}$Ni reaction at various (low) incident energies 
\cite{campo91,campo88,campo98,commara2k}. Also, some of the specific data 
\cite{campo91,campo88} chosen here, and discussed in the following paragraph, 
was not analysed in our earlier work \cite{gupta02}. This system has been of much 
interest from the point of view of the exotic $^{12}$C (or $^{100}$Sn) cluster
radioactivity \cite{commara2k,kumar93,poenaru95,oganess94,gugliel95}, first expected 
as a ground-state decay \cite{kumar93,poenaru95,oganess94,gugliel95}, but now seems 
to have observable yields only for the decay of the corresponding excited compound 
system \cite{commara2k}. Another important point to note is that $^{116}$Ba is mainly 
a positive Q-value (Q$_{out}$) system (negative Q$_{out}$ only for some light 
decay-fragments). Our model is also applied to a negative Q$_{out}$, light compound 
system $^{56}$Ni \cite{gupta03}. Also, the other, both lighter and heavier, isotopes
of barium, $^{112-120,126}$Ba, have been of interest for both the ground-state and 
excited state decay studies \cite{kumar93,royer03}. 

The data chosen to test our model calculations in the following are the 
IMFs production cross sections, measured in (i) the 630 MeV 
$^{58}$Ni+$^{58}$Ni$\rightarrow ^{116}$Ba$^*$ reaction \cite{campo91,campo88}, 
whose kinematical analysis of fragments (the similarity of Gaussian shaped energy 
spectra of fragments in coincidence with $\gamma$ rays, and the flat angular 
distribution of their ground and excited states) revealed that the IMFs 
with 4$\le$Z$\le$9 are consistent with the decay of a compound nucleus 
formed in a complete fusion reaction and that the effective temperature 
of the emitting system, estimated on the basis of a simple statistical
approach for primary fragments, is on the average about 2.2 MeV, which is
about half of the compound nucleus temperature T (=4.42 MeV, based on
Eq. (\ref{eq:10}) below); and (ii) the 348, 371 and 394 MeV 
$^{58}$Ni+$^{58}$Ni reaction \cite{commara2k}, where the pure Hauser-Feshbach 
BUSCO predictions do not match the measured cross sections (the same code 
is used with a reasonable success for the above mentioned data at 630 MeV 
\cite{campo91,campo88}). In other words, so far only the pure Hauser-Feshbach 
analysis is made for both the experiments. Also, like in \cite{campo91,campo88}, 
the total excitation energy (TXE) in the exit channel is observed to be too 
small ($\sim$50 MeV), compared to the available compound nucleus excitation 
energy E$_{CN}^*$ (=96.4 to 130.9 MeV). As is already discussed in our 
earlier work \cite{gupta02}, such data on TXE (or TKE) are best treated in 
a fission or a cluster decay model, and we use here the cluster decay model.

The dynamical cluster-decay model for hot compound systems, a reformulation
of the preformed cluster model (PCM) of Gupta and collaborators
\cite{gupta88,malik89,gupta91,kumar97,gupta99a} for ground-state decays,
is presented in Sec. 2 and its application to the hot $^{116}$Ba$^*$ nucleus
data is made in Sec. 3. The experimental data for IMFs production cross 
sections are taken from Fig. 6 of \cite{campo91}, and from Table 1 or Fig. 2 of 
\cite{commara2k}. It may be noted that any discussion of the emitted light 
particles in this paper is of those predicted here to occur at higher energies, 
and for the $\ell$=0 case only. The (statistical) evaporation of light particles, 
that may occur just before the beginning of the binary decay process of cluster 
emission studied here, is not included in this paper. Finally, a summary and 
discussion of our results is added in Sec. 4.

\section{THE DYNAMICAL CLUSTER DECAY MODEL}
The dynamical cluster-decay model developed here for hot compound nucleus 
is an adaptation of the preformed cluster model (PCM) of Gupta et al.
\cite{gupta88,malik89,gupta91,kumar97,gupta99a} for ground-state decays, 
which itself is based on the well-known quantum mechanical fragmentation 
theory, the QMFT \cite{maruhn74,gupta75,gupta99b}, given for fission and
heavy-ion reactions and used later for the prediction of exotic cluster 
radioactivity \cite{gupta94,sandu80,rose84}. In the preformed cluster model, 
the decay half-life T$_{1\over 2}$, or the decay constant $\lambda $, is 
defined as
\be
\lambda ={{{ln 2}\over {T_{1\over 2}}}}=P_0\nu _0 P, 
\label{eq:1}
\ee
where P$_0$, the preformation probability, refers to the motion in the mass 
asymmetry coordinate $\eta$=(A$_1$-A$_2$)/(A$_1$+A$_2$) and P, the 
penetrability, to the motion in the relative separation coordinate R. This 
means that, like in PCM, we treat the distribution of the IMFs as a dynamical 
collective mass motion of preformed fragments or clusters through the barrier. 
In principle, the $\eta $ and R motions are coupled, but, in view of the 
defining Eq. (\ref{eq:1}) and our earlier works \cite{maruhn74,gupta75,saroha86}, 
these are taken to be decoupled here. The $\nu _0$ is the assault frequency, 
defined below. In terms of the partial waves, the compound nucleus decay or the
fragments production cross section 
\be
\sigma={\pi \over k^2}\sum_{\ell=0}^{\ell_{c}}(2\ell+1)P_0P; 
\qquad 
k=\sqrt{2\mu E_{c.m.}\over {\hbar^2}}
\label{eq:2a}
\ee
with $\mu =[A_1A_2/(A_1+A_2)]m={1\over 4}Am(1-\eta ^2)$ as the reduced mass 
and $\ell_{c}$, the critical (maximum) angular momentum, defined later. m is 
the nucleon mass. This means that $\lambda$ in (\ref{eq:1}) gives the 
cross section for $\ell =0$ case, with a normalization constant $\nu _0$, 
instead of the ${\pi}/{k^2}$ in (\ref{eq:2a}). However, in the present 
calculations, made for $\ell =0$ (s-wave) case, the normalization constant is 
obtained empirically from the experimental data.
 
The preformation probability
\be
P_0={\sqrt {B_{\eta \eta}}}\mid \psi (\eta (A_i))\mid ^2\left ({2/A}\right ),
\label{eq:1a}
\ee
(i=1 or 2), where $\psi (\eta )$ are related to $\psi ^{\nu}(\eta )$,
$\nu$=0,1,2,3... (see Eq.(\ref{eq:9})), the eigen-solutions
of stationary Schr\"odinger equation in $\eta$, at fixed R,
\be
\{ -{{\hbar^2}\over {2\sqrt B_{\eta \eta}}}{\partial \over {\partial
\eta}}{1\over {\sqrt B_{\eta \eta}}}{\partial\over {\partial \eta
}}+V_R(\eta )\} \psi ^{\nu}(\eta ) = E^{\nu} \psi ^{\nu}(\eta ),
\label{eq:2}
\ee
which is solved at R=R$_a$, the first turning point in the inter-nuclear 
potential (see Fig. 1). For the ground-state (T=0) decay 
R$_a$=C$_1$+C$_2$=C$_t$, fixed empirically, since this value of R (instead 
of R=R$_0$, the compound nucleus radius) assimilates to a good extent the 
effects of both the deformations $\beta_i$ of the two fragments and the neck 
formation between them \cite{kumar97}. The C$_i$ are the S\"ussmann central 
radii C$_i$=R$_i$-(b/R$_i$), with the radii 
R$_i$=1.28A$_i^{1/3}$-0.76+0.8A$_i^{-1/3}$ fm and surface thickness 
parameter b=0.99 fm. Thus, by raising the first 
turning point R$_a$ from R$_a$=R$_0$ to R$_a$=C$_t$ or, in general,  
=C$_t$+$\sum \delta R(\beta_i)$, the deformation effects of the two fragments 
(and the neck formation between them) are included here in the scattering 
potential V(R,T=0), within the extended model of Gupta and collaborators for 
both the light and heavy nuclear systems \cite{kumar97,khosla90,gupta97}. The 
neck-length parameter $\sum \delta R(\beta_i)$, introduced at the scission 
configuration, simulates the two-centre nuclear shape parametrization and is 
equivalent to the lowering of the barrier for the case of deformed fragments 
\cite{kumar97}. This method of inclusion of fragment deformations in the
present calculations is quite similar to what has been achieved in both the
saddle-point transition-state model of Sanders \cite{sanders89,sanders91}
and the scission-point Extended Hauser-Feshbach Method of Matsuse and
collaborators \cite{matsuse97}. The work of Ref. \cite{kumar97} also shows
that the alternative of calculating the fragmentation potential $V(\eta)$ 
and scattering potential V(R) for actual deformations of the nuclei is not
practical since the experimental deformation parameters for all the possible
fragments (A$_1$,A$_2$), required for calculating V$(\eta)$, are generally 
not available. Note that the deformation effects of nuclei in the present
calculations are also included via the S\"ussmann central radii C$_i$ and 
that C$_t$ are different for different $\eta$-values and hence C$_t$ is 
C$_t(\eta)$. The temperature effects on radii R$_i$, and hence on C$_i$, 
the neck-length parameter $\delta R(\beta_i)$ and the surface width b are 
discussed in the following. 

For the decay of a hot compound nucleus, we follow the prescription of our
earlier work \cite{gupta02} and define the first turning point
\be
R_a(T)=C_t(\eta ,T)+\Delta R(\eta ,T),
\label{eq:4}
\ee 
which means changing R$_a$ to R$_a+\Delta R$ for the hot compound nucleus. 
Note that $C_t$ and $\Delta R$ are also taken temperature-dependent, as 
defined later. Fig. 1 illustrates the scattering potential and other
quantities of interest. Apparently, $\Delta R$ depends on the total kinetic 
energy TKE(T), and the corresponding potential V(R$_a$) acts like an effective 
Q-value, Q$_{eff}$(T). Since, R$_a$=C$_t(\eta)$ for T=0, the $\Delta R(\eta)$ 
corresponds to the change in TKE at T with respect to its value at T=0 and 
R$_a$=C$_t$, and hence can be estimated exactly for the temperature effects 
included in the scattering potential V(R).

We define Q$_{eff}$(T) in terms of the binding energies of the hot compound 
system and of the cold (T=0) fragments, as
\bea
Q_{eff}(T)&=&B(T)-[B_1(T=0)+B_2(T=0)] \nonumber \\
          &=&TKE(T) \nonumber \\
          &=&V(R_a),
\label{eq:5}
\eea
meaning thereby that the decaying hot compound system, which is at temperature T 
at R=R$_a$, comes out of the barrier and gives in the exit channel two fragments 
that are cold (T=0). These two fragments in the exit channel go to ground state
(T=0) by emitting some light particle(s) and/or $\gamma$-ray(s) of energy (see Fig. 1)
\bea
E_x&=&B(T)-B(0) \nonumber\\
   &=&Q_{eff}(T)-Q_{out}(T=0)\nonumber\\
   &=&TKE(T)-TKE(T=0).
\label{eq:6}
\eea
This can be understood by writing from Eq. (\ref{eq:6}) or (\ref{eq:5}) 
\bea
Q_{eff}(T)&=&TKE(T) \nonumber\\
          &=&Q_{out}(T=0)+E_x \nonumber\\
          &=&TKE(T=0)+E_x,
\label{eq:7}
\eea
which means that the fragments are observed in the ground state with 
Q$_{out}$(T=0) (or =TKE(T=0)) and light particle(s) and/or $\gamma$-rays  
emitted with energy $E_x$. The remaining excitation energy of the decaying 
system is then,
\be
E_{CN}^*-E_x=\mid Q_{out}(T)\mid+TKE(T=0)+TXE(T),
\label{eq:7a}
\ee
where, in the exit-channel the compound nucleus excitation energy 
E$_{CN}^*$ (=$E_{c.m.}+Q_{in}$, the sum of the entrance-channel 
centre-of-mass energy E$_{c.m.}$ and the corresponding Q-value, 
Q$_{in}$=-67.59 MeV for $^{58}$Ni+$^{58}$Ni$\rightarrow ^{116}$Ba)
plus the (positive) Q-value of exit-channel fragments at temperature T,
Q$_{out}(T)$, is distributed into the total excitation energy (TXE) and 
total kinetic energy (TKE) of the two outgoing fragments,
\be
E_{CN}^*+Q_{out}(T)=TKE(T)+TXE(T).
\label{eq:12}
\ee
Eq. (\ref{eq:7a}) again shows that the exit channel fragments are obtained 
with their TKE in the ground-state, i.e. with TKE(T=0). The excitation 
energy TXE(T) in (\ref{eq:7a}) is used in the secondary emission of light 
particles from the primary fragments, which are on the average one 
$\alpha$-particle to two-nucleons \cite{campo88}, but are not treated here. 
Instead, we compare our calculations with the primary pre-secondary-evaporation 
fragments emission data, if available. 

Note that in Eq. (\ref{eq:4}), $\Delta R$ depends on $\eta$. In the following, 
however, based on our earlier works \cite{gupta02,gupta03} and the works of 
other authors \cite{sanders91,matsuse97} (discussed later), we use a constant 
average value $\overline{\Delta R}$, independent of $\eta$, which also takes 
care of the additional $\sum \delta R(\beta_i)$ effects of the deformations of 
fragments and neck formation between them. The $\overline{\Delta R}$ is the 
only parameter of the model, though we shall see later that the structure of 
the calculated (s-wave) mass spectrum is nearly independent of the exact choice 
of this parameter value. Thus, in our calculations, Eq. (\ref{eq:4}) is reduced 
to
\be
R_a(T)=C_t(\eta ,T)+\overline{\Delta R}(T).
\label{eq:8}
\ee
Hence, at temperature T, the preformation factor P$_0$ in Eq. (\ref{eq:1a}) 
is calculated for R=R$_a$(T) of Eq. (\ref{eq:8}), with the temperature 
effects also included in $\psi (\eta )$ through a Boltzmann-like function
\be
\mid \psi \mid ^2 = \sum _{\nu =0}^{\infty}\mid\psi ^{\nu}\mid ^2
exp (-E^{\nu}/T).
\label{eq:9}
\ee
The compound nucleus temperature $T$ (in MeV) is given by
\be
E_{CN}^*=\left ({A/9}\right ){T}^2-T.
\label{eq:10}
\ee

The penetrability $P$ is the WKB tunnelling probability, 
\be
P=exp[-{2\over \hbar}{{\int }_{R_a}^{R_b}\{ 2\mu [V(R)-Q_{eff}]\}
^{1/2} dR}].
\label{eq:3}
\ee
This expression is solved analytically \cite{malik89}. Here R$_b$ is the second 
turning point. For the penetrability P, Eq.(\ref{eq:8}) means that
\be
V(R_a)=V(C_t(\eta ,T)+\overline{\Delta R}(T))=V(R_b)
=Q_{eff}(T)=TKE(T).
\label{eq:13}
\ee

In Eq.(\ref{eq:2}), the fragmentation potential V$_R(\eta )$ at any
temperature T is calculated within the Strutinsky renormalization procedure,
as
\bea
V_R(\eta ,T)&=&\sum_{i=1}^{2}{\Bigl [ V_{LDM}(A_i,Z_i,T)\Bigr ]}+
\sum_{i=1}^{2}{\Bigl [ \delta U_i\Bigr ]}exp(-\frac{{T}^2}{{T_0}^2})
\nonumber\\
&+&{{Z_1Z_2e^2}\over {R(T)}}+V_P(T)+V_{\ell}(T),
\label{eq:14}
\eea
where, as already stated in the Introduction, the T-dependent liquid drop
energy V$_{LDM}(T)$ is that of Ref. \cite{davidson94}, and shell corrections 
$\delta U(T)$ are considered to vanish exponentially, with $T_0=1.5$ MeV. 
The V$_{LDM}(T)$ has the following form, based on the semi-empirical mass
formula of Seeger \cite{seeger61}, 
\bea
V_{LDM}(A,Z,T)&=&\alpha (T)A+\beta (T)A^{2\over 3}+\Bigl (\gamma (T)-{\eta
(T)\over {A^{1\over 3}}}\Bigr )
\nonumber\\
&&\times \Bigl ({{I^2+2\mid I\mid }\over A}\Bigr )+{Z^2\over {r_0(T)A^{1\over
3}}}\Bigl (1-
{0.7636\over {Z^{2\over 3}}}
\nonumber\\
&-&{2.29\over {[r_0(T)A^{1\over 3}]^2}}\Bigr )+\delta (T){f(Z,A)\over {A^{3\over
4}}},
\label{eq:15}
\eea
with
$$I=a_{a}(Z-N),\quad\hbox{$a_{a}=$1,}$$
and, respectively, for even-even, even-odd, and odd-odd nuclei,
$$f(Z,A)=(-1,0,1).$$
Seeger \cite{seeger61} fitted the constants from the ground-state (T=0)
binding energies of some 488 nuclei available at that time (in 1961), and
obtained:
$$\alpha (0)=-16.11 ~\hbox{MeV,}\quad\beta (0)= 20.21 ~\hbox{MeV,}$$
$$\gamma (0)=20.65 ~\hbox{MeV,}\quad\eta (0)=48.00 ~\hbox{MeV,}$$
and from Ref. \cite{benedetti64}, the pairing energy term 
$$\delta (0)=33.0 ~\hbox{MeV.}$$
These constants apparently need to be re-fitted to the large amount of data 
on ground-state binding energy available now \cite{audi95}, particularly
for neutron-rich nuclei. This is partly done in Ref. \cite{gupta03} and is
extended here to nuclei upto Z=56. The constants required to be fitted again
are the bulk constant $\alpha (0)$, working as an overall scaling factor, and
the proton-neutron asymmetry constant $a_a$, controlling the curvature of the
experimental parabola, illustrated in Fig. 2 by the open circles for Z=49 nuclei. 
Note the excellent agreement between the present fits (open circles) and the 
compilation of experimental data of Ref. \cite{audi95} (solid circles). The fits 
are obtained within 1 to 1.5 MeV of the measured binding energies. Table 1 gives
as an example the fitted constants for $1\le Z\le 56$ nuclei, which includes 
the ones in Table 1 of Ref. \cite{gupta03}.
The kind of improvement obtained is further illustrated in Fig. 3 for the
fragmentation potentials, used in the present calculations (compare the newly 
fitted binding energies displayed as open circles with the experimental binding
energies displayed as solid circles \cite{audi95}). The two curves are almost 
identical, particularly for the positions of maxima and minima. 

The temperature dependence of the constants in Eq. (\ref{eq:15}) was 
obtained numerically in Ref.\cite{davidson94} by using the available
experimental information on excited states of 313 nuclei in the mass region
$22\le A\le 250$ for determining the partition function ${\cal{Z}}$(A,Z,T) of
each nucleus in a canonical ensemble (see Ref. \cite{davidson94} for further
details) and making a least squares fit of the excitation energy
$E_{ex}(A,Z,T)=V_{LDM}(A,Z,T)-V_{LDM}(A,Z,0)$ to the ensemble average 
$E_{ex}(A,Z,T)=T^2{\partial\over {\partial T}}ln{\cal{Z}}(A,Z,T).$
The $\alpha (T)$, $\beta (T)$, $\gamma (T)$, $\eta (T)$ and $\delta (T)$ thus
obtained are given in Figure 1 of Ref. \cite{davidson94} for $T\le 4$MeV,
extrapolated linearly for higher temperatures. The $\delta (T)$ is constrained 
to be positive definite at all temperatures, with $\delta (T>2~MeV)=$0. For the 
bulk constant $\alpha (T)$, the authors also give the following empirically 
fitted equation to a Fermi gas model,
$$\alpha (T)=\alpha (0)+{T^2\over 15},$$
which is used here in our calculations with the re-fitted $\alpha (0)$.
Finally, for the radius constant $r_0(T)$ in (\ref{eq:15}), the analytical form, 
taken from Ref. \cite{brack74}, is
\be
r_0(T)=1.07(1+0.01T).
\label{eq:16}
\ee
This means that R$_i$ (=$r_0 A_i^{1\over 3}$), and in turn C$_i$, are
T-dependent.

The shell effects $\delta U$, added to the liquid drop energy in Eq. (\ref{eq:14}), 
are obtained empirically by Myers and Swiatecki \cite{myers66}, for spherical
shapes, as
\be
\delta U=C\left [{{F(N)+F(Z)}\over {({A/2})^{2\over 3}}}-cA^{1\over 3}\right ]
\label{eq:17}
\ee
with 
\be
F(X)={3\over 5} \left ( {{M_i^{5\over 3}-M_{i-1}^{5\over 3}} \over
{M_i-M_{i-1}}}\right )
\left (X-M_{i-1}\right )-{3\over 5}\left (X^{5\over 3}-M_{i-1}^{5\over 3}\right
)
\label{eq:18}
\ee
where X=N or Z, $M_{i-1}<X<M_i$. $M_i$ are the magic numbers 2,8,14 (or 20),
28,50,82,126 and 184 for both neutrons and protons. The constants are C=5.8 MeV 
and c=0.26 MeV. We refer to the use of magic numbers 14 or 20 as MS14 or MS20
parametrization.

The V$_P$ in Eq. (\ref{eq:14}) is the additional attraction due
to the nuclear proximity potential \cite{blocki77}, also taken to be
temperature-dependent, as
\be
V_P(R,T)=4\pi\bar{R}(T)\gamma b(T)\Phi (s,T),
\label{eq:19}
\ee
where $\bar{R}$(T) is the inverse of the root mean square radius of the
Gaussian curvature and $\Phi (s,T)$ is the universal function, independent of
the geometry of the system, given by
\be
\Phi (s,T)=\left \{
\ba{ll}
-{1 \over 2}(s-2.54)^2-0.0852(s-2.54)^3 & \hbox{for} \quad s\le 1.2511 \\
-3.437exp(-{s \over 0.75}) & \hbox{for} \quad s\ge 1.2511
\ea
\right.
\label{eq:20}
\ee
\be
\bar{R}(T)=\frac{C_1(T)C_2(T)}{C_t(T)}.
\label{eq:21}
\ee
Furthermore, $\gamma$ is the specific nuclear surface tension given by
\be
\gamma =0.9517\left[1-1.7826\left(\frac{N-Z}{A} \right)^{2}
\right] MeV fm^{-2}.
\label{eq:22}
\ee
In Eq. (\ref{eq:20}), s(T) (=${{R-C_t(T)}\over b(T)}$) is the overlap
distance, in units of b, between the colliding surfaces. The surface width 
b is \cite{royer92}
\be
b(T)=0.99(1+0.009T^2).
\label{eq:23}
\ee
Thus, in $C_i=R_i-(b/R_i)$, the T-dependence is via both R$_i$ and b.

The same temperature dependence of r$_0(T)$, given by Eq. (\ref{eq:16}), is also
used for the Coulomb potential $E_c(T)=Z_1Z_2e^2/r_0(T)A^{1/3}$ where the
charges $Z_i$ are fixed by minimizing the potential $V_R(\eta ,T)$ in the charge
asymmetry coordinate $\eta_Z$=(Z$_1$-Z$_2$)/(Z$_1$+Z$_2$). The $V_{\ell}$ is the
centrifugal potential. However, like in Refs. \cite{gupta02,gupta03}, we consider 
here only the $\ell$=0 case ($V_{\ell}$=0) since the s-wave is shown to contain 
the gross information about both the measured fragment emission yields and their 
total kinetic energies. For a detailed description, however, the angular momentum 
effects must also be included, which we wish to do in our next publication.

The mass parameters B$_{\eta \eta}(\eta )$, representing the kinetic energy
part in Eq. (\ref{eq:2}), are the smooth classical hydrodynamical masses
\cite{kroeger80}, since we are dealing here with a situation where the
shell effects are almost completely washed out.
In the classical hydrodynamical model, for touching spheres, we use
\be
B_{\eta \eta}={{AmC_t^2(T)}\over 4}{\Biggl ({{v_t(T)(1+\beta (T))}\over
{v_c(T)}}-1 \Biggr )}
\label{eq:24}
\ee 
with
\be
\beta (T)={R_c(T)\over {4C_t(T)}}{\Biggl (2-{R_c(T)\over C_1(T)}-{R_c(T)\over
C_2(T)}\Biggr )}
\label{eq:25}
\ee 
\be
v_c(T)=\pi R_c^2(T) C_t(T),\qquad R_c(T)=0.4C_2(T), 
\label{eq:26}
\ee 
and $v_t=v_1+v_2$, the total conserved volume. Also, $C_2<<C_1$ and 
$R_c(\neq 0)$ is the radius of a cylinder of length C$_t$, whose existence 
allows a homogeneous, radial flow of mass between the two fragments.
Note that here all radial or radial-dependent quantities are
temperature-dependent, with 
$R_i(T)=r_0(T)A_i^{1/3}$ fm in C$_i$ and C$_t$. The above expressions are 
also used for T-independent calculations, but then with
$R_i=1.28A_i^{1/3}-0.76+0.8A_i^{-1/3}$ fm.

In Eq.(\ref{eq:1}), the assault frequency $\nu _0$ is given simply as
\be
\nu _0={{(2E_2/\mu )^{1/2}}\over R_0},
\label{eq:27}
\ee
with the kinetic energy of the lighter fragment $E_2=(A_1/A) Q_{eff}$, where 
Q$_{eff}$ is shared between the two fragments as inverse of their masses.
However, for the calculations of s-wave cross sections, instead of $\nu _0$, 
we use an empirically determined normalization constant.

Finally, the temperature-dependent scattering potential $V_{\eta}(R,T)$,
normalized to the exit channel binding energy, is
\be
V_{\eta}(R,T)={{Z_1Z_2e^2}\over {R(T)}}+V_P(T), 
\label{eq:28}
\ee
with $R_i(T)=r_0(T)A_i^{1/3}$ fm. This is illustrated in Fig. 1. For 
T-independent case $R_i=1.28A_i^{1/3}-0.76+0.8A_i^{-1/3}$ fm. Eq. (\ref{eq:28}) 
means that for fixed $\eta$, the energies are measured w.r.t. $B_1(T)+B_2(T)$,
and the fragments go to ground state ($T\rightarrow 0$) only via the emission
of particle(s) or $\gamma$-ray(s) of energy $E_x$.

\section{CALCULATIONS}

In this section, we present our calculations in such a way that the role of 
temperature is emphasised in the different terms of the potential. For this
purpose, we first look at our calculations for the temperature effects only 
in liquid drop energy and shell corrections, i.e. for use of V$_{LDM}$(T) and 
$\delta U$(T), but T-independent E$_c$ and V$_P$. Note that this is already a 
first improvement over our earlier calculation in \cite{gupta02}, where only 
shell corrections were taken to be T-dependent. Then, as a second and final 
improvement, we include the T-dependence in all terms of the potential, i.e. 
we use V$_{LDM}$(T), $\delta U$(T), E$_c$(T) and V$_P$(T). The resulting 
fragmentation potentials for $^{116}$Ba are given in the following two 
subsections, respectively, for each case. Only s-wave ($\ell =0$) solutions 
are studied here in the following.

\subsection{Temperature-dependence only in liquid drop energy and shell
correction}
Figure 4 gives our calculated fragmentation potentials V$(\eta )$ for cold
(T=0) as well as hot $^{116}$Ba$^*$ nucleus at various T-values referring to
compound nucleus excitation energies E$_{CN}^*$ of the experiments of Refs.
\cite{campo91,commara2k}. Comparing with our earlier calculation
\cite{gupta02}, here the liquid drop potential V$_{LDM}$ is also T-dependent.
The R$_a$-value for each T-value is chosen as follows: At T=0,
R$_a$=C$_t$, as is already discussed above. For hot compound systems,
we fix R$_a$=C$_t+\overline{\Delta R}$ with $\overline{\Delta R}=$ 0.226, 0.252, 
0.281, and 0.466 fm (chosen arbitrarily), respectively, for T=2.91, 3.06, 3.20, 
and 4.42 MeV, which correspond to the four experimental values of 
E$_{c.m.}$=174, 185.5, 197, and 315 MeV of Refs. \cite{commara2k} and 
\cite{campo91}. We notice that, as far as the fine structure is concerned, the 
calculations are nearly independent of the exact choice of $\overline{\Delta R}$-value, 
which is explicitly illustrated in Fig. 5 for T=4.42 MeV. This observation has already 
been made in our earlier studies on QMFT \cite{gupta99a,gupta99b}. As a consequence, 
the same is true for the calculated yields shown below in Figs. 8 and 9. Thus, for a 
given E$_{c.m.}$, $\overline{\Delta R}$ is fixed for all the fragments; on the other 
hand $\overline{\Delta R}$ was fitted for each fragment separately in our earlier work 
\cite{gupta02}, as well as in the statistical model calculations of other authors 
\cite{sanders91,matsuse97}. Note, the $\overline{\Delta R}$ in our model is the 
same as the neck length parameter $d$ in Refs. \cite{sanders91,matsuse97}.

We notice in Fig. 4 that, due to the T-dependence of V$_{LDM}$, the
fragmentation potentials show large differences as compared to our earlier
calculations in Ref. \cite{gupta02}. First of all, the $\alpha$-like
structure, present at all T values in our earlier calculations, including
the extreme case of T=4.42 MeV where $\delta $U(T) is reduced almost to zero,
is now washed out, even at the lowest temperature. The potential energy
surfaces are now smooth for the heavier fragments at all T values, and new
non-$\alpha$ fragments appear in the lighter mass region. In other words, with
increased T, not only the shell structure effects go to zero but also the
preference for $\alpha$-like structure, due to the Wigner term, vanishes. Also, 
the light particle (A$\le$3) structure is found to change with T. Thus, with T
added, the significant structure in V$(\eta )$ seems to remain  only for a
small window of light fragments having masses less than $\sim$20, and we
refer to it as the "IMF window" in the following.

Figure 6 shows the preformation factor P$_0$ for the IMF window only, calculated
for the four chosen experimental temperatures. We notice that the formation
yields increase as T increases and that some of the fragments are strongly
preformed compared to their neighbours. Also, the yields for fragments with
masses A$_2>$16 are very small, with $^{12}$C being better preformed at higher
energies. The lighter particles (A$\le$3), other than the statistically evaporated
ones (not considered here), are also equally strongly preformed and a neutron
(n) at lower energies changes to a proton ($^1$H) at higher energies. However, the 
final decay constant (equivalently, the IMF s-wave production cross section) is a
combined effect of P$_0$ and P, the $\nu _0$ being almost constant. The result
of this calculation is presented in Fig. 7 for all the strongly preformed and
observed IMFs (Z$<$9, A$\le$18) at the four experimental energies. Some of the
light particle emissions (n and $^2$H) are not shown here since their
Q$_{eff}$ values are negative (or nearly zero) for all the T and
R$_a$ values considered here (the dependence of the decay constant
$\lambda$ on R$_a$ is already studied in our earlier work \cite{gupta02}; see
also Fig. 9 below). For $^1$H, Q$_{eff}$ is positive only for the two higher
energies corresponding to T=3.20 and 4.42 MeV.

The role of the penetrability P is evident in Fig. 7, since some of the
strongly preformed fragments, like $^{4}$He and $^{6}$Li, are now shown as
less favoured decays (smaller decay constants). $^4$He, which has the largest
preformation factor and decay constant for the ground-state decay (see Figs. 3
and 4 in Ref. \cite{gupta02}), lies here lowest for the hot $^{116}$Ba$^*$. 
On the other hand, the other two light particle ($^1$H and $^3$He) decays become 
the most probable ones at all T values here, and are kind of competing with the 
so-called statistically evaporated particles. Also, the non-$\alpha$ like decays,
like $^{10}$B and $^{14}$N, with relatively low P$_0$'s, are now the most
predominant ones. In other words, with the T-dependence of V$_{LDM}$ included,
the exclusive preference for $\alpha$-like IMFs seems lost, the decay rates
for non-$\alpha$ like decays are higher, and the light particle production rates
are the highest. However, the $^{12}$C decay is still striking since it is the
best $\alpha$-nucleus decay even now, particularly at lower energies (inspite
of the lower preformation factor P$_0$ at these energies). At higher energies,
the lighter $\alpha$-like fragment $^{8}$Be competes with it, though we know that
$^8$Be is unbound in the ground state. Note that the charge number Z changes with 
T for some fragments (here for the mass 17 fragment) and hence at higher T's the 
measurements of charge distributions become equally important \cite{charity88,campo98}.
The sudden increase in the production cross section ($\lambda$-value) for $^{17}$F at 
T$\ge$3.2 MeV is due to its enhanced penetrability P, the Q-value effect. (The same effect 
is observed later in Fig. 13 for T-dependence in all terms of the potential, but at
higher temperature T=4.42 MeV only).

Fig. 8, upper panel, shows the IMF production cross sections $\sigma$ for
Z$_2\le$9, measured as compound nucleus decays at E$_{c.m.}$=315 MeV
\cite{campo91}, compared with our calculated s-wave production cross sections
i.e. $\lambda$-values for $\overline{\Delta R}$=0.466 {\it normalized} to the 
experimental data for Z$_2$=6 fragment decay. The calculated yields are for the 
energetically favoured, most probable fragments in both the mass and charge 
numbers (see Figs. 4 and 6) and hence the yields for the same Z$_2$ but different 
masses are summed up. For example, the fragments contributing to the Z$_2$=6 fragment 
yield are the $^{11}$C, $^{12}$C, and $^{13}$C fragments, and the same to the Z$_2$=7 
fragment yield are the $^{14}$N and $^{15}$N fragments, etc. For Z$_2$=4, however,
we should be cautious in making this comparison between the experiment and calculation
since $^8$Be, being unbound as a ground state, may not be measured, but we have included 
it in our summed yields for Z$_2$=4 (the other component is $^9$Be, with a rather small 
yield $\sim 10^{-4}$). The lower panel shows the same for another normalization of our 
calculations to the Z$_2$=7 fragment data. Furthermore, Fig. 9 presents the same 
comparison (only for normalization to the Z$_2$=6 fragment data) for another two 
$\overline{\Delta R}$ values. Apparently, the calculations compare reasonably well with 
experiment and are almost independent of the (empirical) normalization constant as well 
as the exact choice of $\overline{\Delta R}$ value. For the heavier fragments (Z$_2\ge$8), 
the calculated cross sections are rather small firstly because we are considering here 
the contribution of only $\ell =0$ term and secondly because the temperature effects are 
included only in V$_{LDM}$ and $\delta$U. Also, in experiments \cite{campo91,campo88} the 
contributions from other sources such as the peripheral collisions like projectile 
splitting or deep-inelastic-like mechanism are difficult to be separated out for the 
heavy fragments. Thus, in view of the fact that we are dealing here with only the 
$\ell =0$ case and that the temperature effects are not fully included in the potential, 
the comparisons between the theory and experiments in Figs. 8 and 9 could be considered 
rather satisfactory.

A comparison of our calculations with the experimental data at the other three lower 
energies \cite{commara2k} is shown in Fig. 10 for only the four heavy residues $^{99}$Cd,
$^{100}$In, $^{101}$Sn and $^{102}$In that have been actually measured at each of these 
three energies, as reported in Ref. \cite{commara2k} (see Fig. 2 and Table 1 of 
\cite{commara2k}). In this case, the calculations have been normalized arbitrarily to 
the A$_1$=102 fragment. Some calculated fragments show a change of fragment charge 
number since we have considered here only the energetically favoured, most probable 
fragment in both mass and charge numbers which could be different from the one actually 
measured. The comparison shows that the predicted yields are in reasonable agreement 
with the experimental data.

\subsection{Temperature-dependence of the full potential}
Figure 11 gives the calculated fragmentation potentials V$(\eta ,T)$ for
T-dependence added in all the terms of the potential (Eq. (\ref{eq:14}), for
$\ell =0$). The T and $\overline{\Delta R}$ values are the same as in Fig. 4
for no T-effects in E$_c$ and V$_P$. (The graph for T=0 is again shown here
for comparisons). We notice that there is hardly any sizeable effect due to the 
additional T-dependences in E$_c$ and V$_P$, when Fig. 11 is compared with Fig. 4, 
at least for the structure and positions of minima, especially at lower T-values. 
At higher T-values, there are some very small changes, particularly for very light
fragments (A$_2\le$2). It is known \cite{saroha85} that V$_P$ does not contribute 
towards the structure and positions of minima in the fragmentation potential V$(\eta )$
and that E$_c$ is a smoothly increasing function of A$_2$. However, the mass 
parameters B$_{\eta \eta}$, though smooth, are now T-dependent, and this behaviour
seems to enhance the formation probability P$_0$, presented in Fig. 12. 
The enhancement is by about one order of magnitude (compare Fig. 12 with Fig. 6 
without T-dependence in E$_c$, V$_P$, and B$_{\eta \eta}$). As a consequence, we
note that in the formation yield of $^{12}$C, relative to its neighbouring fragments, 
the small peaked structure present in Fig. 6 is now absent in Fig. 12.

Figure 13 presents the result of our calculation for the decay constant $\lambda$
(equivalently, the s-wave production cross section) as a function of compound 
nucleus temperature corresponding to four energies (174, 185.5, 197, and 315 MeV) of 
the two selected experimental data sets of Refs. \cite{campo91,commara2k}. We notice 
important differences with respect to the case of no temperature in E$_c$ and V$_P$ 
(compare Fig. 13 with Fig. 7). Here, for light particles, instead of only $^1$H and 
$^4$He at higher temperatures, we get $^2$H, $^3$H and $^4$He at all the temperatures 
which is favoured by the experimental data (for the neutron, the Q$_{eff}$ is 
still negative). Note that here these light fragments are not the
statistically treated prompt particles, though the calculated yields are shown
to be large (calculations made so far are only for the $\ell =0$ case).
Interesting enough, $^{12}$C decay is still the most favoured $\alpha$-nucleus
decay, though other non-$\alpha$ fragments are shown to compete. Also, some
fragments having smaller cross sections at lower energies are shown to have
enhanced cross sections at the higher energy. 

Finally, a comparison of our calculated s-wave cross sections with the measured 
cross sections is made in Figs. 14 and 15. Fig. 14 shows this comparison for the 
E$_{c.m.}$=315 MeV data \cite{campo91} with our calculated decay constants 
(s-wave cross sections) normalized to Z$_2$=6 fragment data. Similarly, Fig. 15 
gives a comparison of our calculations with the measured cross sections for the 
three lower energies of Ref. \cite{commara2k}. Here the data are obtained 
for only the heavy fragments and we have normalized our calculations to the 
mass 102 fragment. The comparisons between the calculations and data are rather 
good at all energies, and better than for the earlier case displayed in Figs. 8, 9 
and 10. The improvement demonstrates the need of including temperature dependence in 
all the terms of the potential. 

\section{Summary and discussion}
Summarizing, first the use of T-dependent V$_{LDM}$ is studied for the cluster 
decay process proposed for the IMFs emission \cite{gupta02} induced in heavy-ion
collisions at low bombarding energies (E$_{c.m.}<$15 MeV/A). The model is extended and
applied to $^{58}$Ni+$^{58}$Ni$\rightarrow ^{116}$Ba$^*$ at various centre-of-mass 
energies from 174 to 315 MeV. Interesting and important enough, with the use of a
T-dependent V$_{LDM}$ (for $\ell$=0 case only), the explicit preference for 
$\alpha$-like IMFs (or clusters) is lost, though the $^{12}$C (or $^{104}$Sn) 
decay (first expected from the ground state decay of $^{116}$Ba 
\cite{kumar93,poenaru95}) is still found to be the most preferred $\alpha$-like 
fragment. Also, the non-$\alpha$ IMFs and some light particles with Z$<$2 are 
shown to be produced with larger yields, without the inclusion of any statistical 
evaporation process. With T-dependence included in all the terms of the potential, 
the production yields are enhanced and comparisons with data are improved.
Further improvements are expected with the inclusion of contributions from all
$\ell$-waves, which is being worked out.

Also, the TKEs are measurable quantities, but the same are available for the 
$^{58}$Ni+$^{58}$Ni reaction only for E$_{c.m.}$=315 MeV \cite{campo91} and at
only one angle and four charges (without the mass identification of IMFs). In our
model, as already defined above and in Fig. 1, the angle-integrated total kinetic
energies are given by TKE=V(R$_a$)=V(C$_t+\Delta$R)=Q$_{eff}$. Since the calculated 
production yields are almost independent of the (arbitrarily) chosen average 
$\overline{\Delta R}$ values, only the relative variation and not the absolute 
magnitude of TKEs could be obtained at present in this model. Furthermore, the angular 
momentum contribution (not included as yet) is a must for TKEs calculations. Such type 
of calculations have already been performed \cite{gupta03} for the lighter system 
$^{32}$S+$^{24}$Mg$\rightarrow ^{56}$Ni where the angle integrated TKEs data were 
available \cite{sanders89}. It will be interesting to apply similar considerations 
to $^{116}$Ba$^*$ when similar data on TKEs become available. Also, it would be of 
interest if the statistical fission model, as introduced in the GEMINI code 
\cite{charity88} could also be applied to the $^{58}$Ni+$^{58}$Ni reaction and 
its predictions compared with the present cluster decay model results as well as 
to the pure Hauser-Feshbach BUSCO results. Furthermore, a similar comparison using
the alternate picture as generalized liquid-drop model \cite{royer01} or the  
so-called unified fission models for cluster decay studies \cite{gupta94} is called for
since these models do not introduce any preformation factor.
\\

\section{ACKNOWLEDGMENTS}
This work is supported in parts by the Council of Scientific and Industrial
Research (CSIR) and the Department of Science and Technology (DST), New Delhi,
India. The IN2P3/Universit\'e Louis Pasteur and IReS Strasbourg, France and the 
VW-Stiftung, Germany are also acknowledged for financial supports.


\newpage
\par\noindent
{\bf FIGURE CAPTIONS}
\begin{description}
\item{Fig. 1.} The s-wave scattering potential, illustrated for
$^{116}$Ba$^*\rightarrow ^{12}$C+$^{104}$Sn, showing the effect of temperature
for its inclusion in all terms of the potential (solid line), i.e.
V(R,T)=E$_c$(T)+V$_P$(T), with the binding energy
B(T)=V$_{LDM}$(T)+$\delta$U(T). The decay path is shown to begin at R$_a$ for 
the T$\ne$0 case, along with other quantities that are discussed in the text. 
\item{Fig. 2.} The mass excess $\Delta M$ (=M$_A$-A=NM$_n$+ZM$_p$+B(Z,N)-A in
MeV) as a function of neutron number N for Z=49 nuclei, calculated by using the
1995 experimental data (solid circles) \cite{audi95}, with newly fitted constants 
(open circles) and with the 1961 Seeger's constants \cite{seeger61} (open squares).
\item{Fig. 3.} The fragmentation potential for $^{116}$Ba at T=0, using the
experimental binding energies (solid circles) \cite{audi95} and the empirical 
binding energies (open circles) with the new constants given in Table 1.
\item{Fig. 4.} The fragmentation potential V($\eta$,R,T) for the hot
$^{116}$Ba$^*$ nucleus, calculated at the ground-state (T=0, R$_a$=C$_t$) and
at the various nuclear temperatures T, which correspond to the bombarding energies 
of the measurements of Refs. \cite{campo91,commara2k}, and at the indicated 
R$_a$-values. The T-dependence is included only in V$_{LDM}$ and $\delta U$, i.e.
V($\eta$,R,T)=V$_{LDM}$(T)+$\delta$U(T)+E$_c$+V$_P$.
The thin vertical lines are drawn to mark the exclusive $\alpha$-like
structure present in T=0 case only.
\item{Fig. 5.} Same as for Fig. 4, but for only T=4.42 MeV and at different
$\overline{\Delta R}$ values.
\item{Fig. 6.} The IMFs preformation probability P$_0$ for the hot
$^{116}$Ba$^*$ nucleus, calculated for the four experimental T-values corresponding 
to E$_{c.m.}$=174, 185.5, 197, and 315 MeV or equivalently T=2.91, 3.06, 3.20 and
4.42 MeV, respectively, using the fragmentation potentials displayed in Fig. 4.
\item{Fig. 7.} The cluster decay constant $\lambda$(s$^{-1}$) as a function of
nuclear temperature T for emissions of the IMFs and light particles from the hot
$^{116}$Ba$^*$ nucleus, calculated at the four experimental T and R$_a$
values used in Fig. 4 that gives the relevant fragmentation potentials. The dashed
line shows the change of charge number of the cluster.
As for the experimental data of Refs. \cite{campo91,campo88}, only the fragments with 
Z$\le$9 are found to be of a compound nucleus origin.
\item{Fig. 8.} The calculated (s-wave) and measured cross sections $\sigma$(mb) 
for all the IMFs produced in decay of $^{116}$Ba$^*$ at E$_{c.m.}$=315 MeV. The
experimental data are taken from the Fig. 6 of Ref. \cite{campo91}. Since in 
experiments \cite{campo91} only the charges of fragments are measured, in calculations 
we have summed up the yields for most probable fragments with the same charge but 
different masses. The calculations are made for $\overline{\Delta R}$=0.466 fm and 
are normalized to experimental data of Z$_2$=6 fragment (upper panel) and Z$_2$=7 
fragment (lower panel).
\item{Fig. 9.} Same as for Fig. 8 (upper panel), but for 
$\overline{\Delta R}$=0.578 fm (upper panel) and 0.8 fm (lower panel).
\item{Fig. 10.} Same as for Fig. 8, but for E$_{c.m.}$=174, 185.5 and 197 
MeV \cite{commara2k} and heavy fragments, and for normalization with respect to 
A$_1$=102 mass fragment. The experimental data are from Table 1 or Fig. 2 of 
Ref. \cite{commara2k}. Note that the calculated and measured Z-values of some 
fragments are different since in calculations we have considered only the energically
favoured fragment with most probable charge and mass.
\item{Fig. 11.} Same as for Fig. 4, but for the T-dependence included in 
all terms of the potential, i.e.
V($\eta$,R,T)=V$_{LDM}$(T)+$\delta$U(T)+E$_c$(T)+V$_P$(T). Here the potential
for only the light fragments is shown, since it is symmetrical with respect to
A$_2$=58 (or $\eta$=0).
\item{Fig. 12.} Same as for Fig. 6, but for the potentials given in Fig. 11,
i.e. for T-dependence included in all terms of the potential.
\item{Fig. 13.} Same as for Fig. 7, but for the potentials given in Fig. 11,
i.e. for T-dependence included in all terms of the potential. The mass A$_2$=1
fragment (n, in this case) is not shown because Q$_{eff}$ is negative for 
the chosen parameters of the calculations.
\item{Fig. 14.} Same as for Fig. 8 (upper panel), but for the potentials given
in Fig. 11, i.e. for T-dependence included in all terms of the potential.
\item{Fig. 15.} Same as for Fig. 10, but for the potentials given in Fig. 11,
i.e. for T-dependence included in all terms of the potential.
\end{description}

\newpage
\begin{table}{Table 1: Re-fitted bulk and asymmetry constants of Seeger's
liquid drop energy for $1\le Z\le 56$.} \\ 
\begin{tabular}{|c|c|c|c||c|c|c|c|}\hline
Z & N & $\alpha (0)$ & a$_a$ & Z & N & $\alpha (0)$ & a$_a$  \\ \hline
1&    2       &-15.85&0.10&6&    4       &-15.70&0.10 \\ \hline
 &    3       &-16.95&0.12& &   5,7      &-16.50&0.10 \\ \hline
 &    4       &-13.00&0.05& &    6       &-16.65&0.10 \\ \hline
 &    5       &-13.70&0.12& &    8       &-15.90&0.10 \\ \hline
2&    1       &-15.50&0.10& &    9       &-15.70&0.10 \\ \hline
 &    2       &-16.00&0.10& &   10       &-15.10&0.10 \\ \hline
 &    3       &-16.80&0.30& &   11       &-14.80&0.10 \\ \hline
 &   4,5      &-14.20&0.30& &12,13,15,16 &-15.00&0.80 \\ \hline
 &    6       &-13.50&0.10& &   14       &-14.85&0.80 \\ \hline
 &   7,8      &-13.00&0.10&7&    3       &-14.30&0.20 \\ \hline
3&  1,2,4,5   &-16.60&0.10& &    4       &-15.20&0.50 \\ \hline
 &    3       &-16.98&0.98& &    5       &-16.20&0.80 \\ \hline
 &    6       &-13.80&0.98& &    6       &-16.55&0.80 \\ \hline
 &    7       &-14.30&0.40& &    7       &-16.80&0.80 \\ \hline
 &   8,9      &-13.20&0.10& &    8       &-16.30&0.80 \\ \hline
4&    1       &-13.00&0.01& &    9       &-16.20&0.80 \\ \hline
 &    2       &-14.50&0.10& &  10,11     &-15.90&0.94 \\ \hline
 &    3       &-16.20&0.80& &   12       &-15.75&0.94 \\ \hline
 &    4       &-16.98&0.98& &   13       &-15.80&0.94 \\ \hline
 &    5       &-16.70&0.60& &   14       &-15.65&0.94 \\ \hline
 &    6       &-15.50&0.80& &   15       &-15.90&0.94 \\ \hline
 &    7       &-15.30&0.50& &   16       &-16.00&0.94 \\ \hline
 &    8       &-14.30&0.10& &   17       &-16.10&0.93 \\ \hline
 &    9       &-14.00&0.10&8&    4       &-14.00&0.94 \\ \hline
 &   10       &-13.30&0.01& &    5       &-15.25&0.94 \\ \hline
5&    2       &-14.60&0.10& &    6       &-15.90&0.94 \\ \hline
 &    3       &-16.50&0.10& &    7       &-16.35&0.94 \\ \hline
 &    4       &-16.60&0.60& &    8       &-16.20&0.94 \\ \hline
 &    5       &-16.99&0.10& &    9       &-16.18&0.94 \\ \hline
 &    6       &-16.60&0.60& &   10       &-15.95&0.94 \\ \hline
 &    7       &-16.30&0.10& &   11       &-15.93&0.94 \\ \hline
 &    8       &-15.35&0.10& & 12,14      &-15.85&0.94 \\ \hline
 &    9       &-15.10&0.10& &   13       &-15.90&0.94 \\ \hline
 &   10       &-14.45&0.10& &   15       &-16.10&0.94 \\ \hline
 &   11       &-14.10&0.10& &   16       &-16.15&0.90 \\ \hline
 &   12       &-13.45&0.10& &   17       &-16.30&0.92 \\ \hline
 &   13       &-13.10&0.10& &   18       &-16.11&0.92 \\ \hline
 &   14       &-13.00&0.40&9&    5       &-15.25&0.90 \\ \hline
6&    2       &-13.00&0.10& &    6       &-15.90&0.90 \\ \hline
 &    3       &-13.85&0.80& &    7       &-16.28&0.90 \\ \hline
\end{tabular}
\end{table}
\begin{table}{Table 1: Continued.....} \\ \\
\begin{tabular}{|c|c|c|c||c|c|c|c|}\hline
Z & N & $\alpha (0)$ & a$_a$ & Z & N & $\alpha (0)$ & a$_a$  \\ \hline
9 &      9          &-16.30 & 0.90 &21 & 15-23,31-38     &-16.42 & 0.77\\ \hline
  &     10          &-16.15 & 0.90 &   &    24-30        &-16.38 & 0.78\\ \hline
  &8,11,17,19,20   &-16.20 & 0.90 &22 &    16-39        &-16.42 & 0.77\\ \hline
  &     12          &-16.01 & 0.90 &23 &    17-40        &-16.42 & 0.77\\ \hline
  &     13          &-16.05 & 0.90 &24 &    18-25        &-16.45 & 0.77\\ \hline
  &     14          &-15.95 & 0.90 &   &    26-41        &-16.42 & 0.77\\ \hline
  &  15,16,18       &-16.11 & 0.90 &25 &    19-26        &-16.46 & 0.77\\ \hline
10&      6          &-15.25 & 0.50 &   &    27-42        &-16.42 & 0.77\\ \hline
  &      7          &-15.70 & 0.50 &26 &    19-43        &-16.46 & 0.77\\ \hline
  &      8          &-15.90 & 0.90 &27 &    21-28        &-16.48 & 0.77\\ \hline
  &     13          &-15.95 & 0.50 &   &    29-45        &-16.46 & 0.77\\ \hline
  &     14          &-15.70 & 0.50 &28 &    22-48        &-16.48 & 0.77\\ \hline
  & 9-12,15-22      &-16.16 & 0.88 &29 &  23-35,49-57    &-16.50 & 0.76\\ \hline
11&      7          &-15.55 & 0.50 &   &    36-48        &-16.48 & 0.77\\ \hline
  &      8          &-15.80 & 0.50 &30 &    24-36        &-16.50 & 0.76\\ \hline
  &     14          &-15.95 & 0.50 &   &    37-52        &-16.48 & 0.70\\ \hline
  & 9-13,15-24      &-16.20 & 0.86 &31 &    25-53        &-16.50 & 0.75\\ \hline
12&   8-10          &-16.11 & 0.90 &32 &    26-34        &-16.57 & 0.75\\ \hline
  &   11-25         &-16.20 & 0.86 &   &  35-40,53,54    &-16.52 & 0.75\\ \hline
13&   8-10          &-16.11 & 0.90 &   &    41-52        &-16.50 & 0.75\\ \hline
  &   11-26         &-16.22 & 0.84 &33 &    27-36        &-16.57 & 0.75\\ \hline
14&   8-12          &-16.11 & 0.90 &   &    37-56        &-16.52 & 0.75\\ \hline
  & 13-20,27,28     &-16.28 & 0.84 &34 &    31-38        &-16.57 & 0.75\\ \hline
  &    21-26        &-16.22 & 0.84 &   &  39-41,53-58    &-16.54 & 0.75\\ \hline
15& 9-13,20-31      &-16.30 & 0.82 &   &    42-52        &-16.52 & 0.75\\ \hline
  &    14-19        &-16.36 & 0.78 &35 &  32-41,53-59    &-16.56 & 0.75\\ \hline
16& 10-14,21-28     &-16.30 & 0.82 &   &    42-52        &-16.54 & 0.75\\ \hline
  &    15-20        &-16.40 & 0.78 &36 &    33-41        &-16.60 & 0.75\\ \hline
  &    29-33        &-16.32 & 0.80 &   &    42-61        &-16.56 & 0.75\\ \hline
17&11-14,20,21,29-34&-16.36 & 0.78 &37 &  34-41,59-65    &-16.63 & 0.75\\ \hline
  &    15-19        &-16.45 & 0.78 &   &    42-58        &-16.58 & 0.75\\ \hline
  &    22-28        &-16.32 & 0.82 &38 &    35-42        &-16.63 & 0.75\\ \hline
18&12-14,21,22,31-35&-16.36 & 0.78 &   &    43-66        &-16.59 & 0.75\\ \hline
  &    15-20        &-16.45 & 0.78 &39 &  38-43,59-67    &-16.63 & 0.75\\ \hline
  &    23-30        &-16.32 & 0.78 &   &    44-58        &-16.61 & 0.75\\ \hline
19&13,14,22,23,30-36&-16.38 & 0.78 &40 &    39-42        &-16.67 & 0.75\\ \hline
  &    15-21        &-16.44 & 0.78 &   &    43-68        &-16.63 & 0.75\\ \hline
  &    24-29        &-16.36 & 0.80 &41 &    40-44        &-16.66 & 0.75 \\ \hline
20& 14,15,22-37     &-16.38 & 0.78 &   &    45-69        &-16.63 & 0.75 \\ \hline
  &    16-21        &-16.48 & 0.78 &42 &  41-45,65-71    &-16.68 & 0.75 \\ \hline
\end{tabular}
\end{table}
\begin{table}{Table 1: Continued.....} \\ \\
\begin{tabular}{|c|c|c|c||c|c|c|c|}\hline
Z & N & $\alpha (0)$ & a$_a$ & Z & N & $\alpha (0)$ & a$_a$  \\ \hline
 42 & 46-64       & -16.64 & 0.75 & 50 & 58-63       & -16.72 & 0.74 \\ \hline
 43 & 42-47,67-72 & -16.68 & 0.75 &    & 64-74       & -16.69 & 0.735 \\ \hline
    & 48-66       & -16.64 & 0.75 &    & 75-87       & -16.67 & 0.72  \\ \hline
 44 & 43-46,69-74 & -16.70 & 0.75 & 51 & 52-61       & -16.75 & 0.75  \\ \hline
    & 47-49       & -16.68 & 0.75 &    & 62-67,83-88 & -16.70 & 0.72  \\ \hline
    & 50-68       & -16.66 & 0.75 &    & 68-82       & -16.67 & 0.72  \\ \hline
 45 & 44-49,69-74 & -16.70 & 0.75 & 52 & 54-58,85-89 & -16.79 & 0.74  \\ \hline
    & 50-68       & -16.67 & 0.75 &    & 59-64,77-84 & -16.75 & 0.74  \\ \hline
 46 & 45-49       & -16.72 & 0.75 &    & 65-76       & -16.73 & 0.74  \\ \hline
    & 50-53,69-73 & -16.70 & 0.75 & 53 & 55-59,83-91 & -16.79 & 0.74  \\ \hline
    & 54-68       & -16.68 & 0.75 &    & 60-82       & -16.75 & 0.74  \\ \hline
 47 & 47-53,77-80 & -16.72 & 0.74 & 54 & 56-59       & -16.83 & 0.74  \\ \hline
    & 54-57       & -16.70 & 0.75 &    & 60-63,83-93 & -16.79 & 0.74  \\ \hline
    & 58-76       & -16.68 & 0.74 &    & 64-82       & -16.77 & 0.74  \\ \hline
 48 & 48-52       & -16.75 & 0.74 & 55 & 57-62       & -16.83 & 0.75  \\ \hline
    & 53-55,77-82 & -16.72 & 0.74 &    & 63-67,81-84 & -16.79 & 0.74  \\ \hline
    & 56-76       & -16.69 & 0.74 &    & 68-80       & -16.77 & 0.74  \\ \hline
 49 & 49-53,75-76 & -16.75 & 0.75 &    & 85-96       & -16.77 & 0.725 \\ \hline
    & 54-59       & -16.72 & 0.74 & 56 & 58-61       & -16.85 & 0.74  \\ \hline
    & 60-74       & -16.69 & 0.74 &    & 62,63,89-97 & -16.84 & 0.74  \\ \hline
    & 77-85       & -16.67 & 0.72 &    & 64-76       & -16.82 & 0.76  \\ \hline
 50 & 50-57       & -16.76 & 0.75 &    & 77-88       & -16.77 & 0.725 \\ \hline
\end{tabular}
\end{table}

\end{document}